\documentclass{article}
\usepackage{graphicx} 
\usepackage{amsmath}
\usepackage{tabularx}
\usepackage{caption}
\usepackage{subcaption}
\usepackage{cite} 

\usepackage[margin=1in]{geometry}
\title{Lie symmetry, Painlev\'e analysis and Conservation laws for (1+2)-Dimensional Kudryashov-Sinelshchikov (KS) equation}
\author{
	Apeksha Patil$^{1}$,
	Amlan Kanti Halder$^{2}$,
	Rajeswari Seshadri$^{3}$\\[1ex]
	$^{1,3}$Department of Mathematics, Pondicherry University, Puducherry, India\\
	$^{2}$School of Sciences, Woxsen University, Hyderabad, India\\
$^{1}$	\texttt{patilapeksha2499@gmail.com}\\
$^{2}$	\texttt{amlanhalder1@gmail.com}\\
Corresponding Author: $^{3}$\texttt{seshadrirajeswari@pondiuni.ac.in}
}

\date{}
\begin{document} 
	\maketitle
	\section*{Abstract}
	The wave propagation of pressures in liquids that contain gas bubbles are an important concern in fluid dynamics and mathematical physics. The Kudryashov Sinelshchikov equation offers a useful mathematical framework in the study of nonlinear wave motion in bubbly liquids with reference to the effects of viscosity and heat exchange between liquid and gaseous phases. This paper examines the dimensional reduced (1 + 2)-dimensional Kudryashov Sinelshchikov equation, a fourth-order nonlinear partial differential equation. Analyzing the Lie symmetry, an infinite dimension Lie algebra is obtained because of the presence of arbitrary functions. By applying the commutative relation between these vector fields and choosing the specific forms for the arbitrary functions, helps the governing PDE to reduce to fourth order ODEs. The reduced equations are then investigated using the Painlevé analysis to give solutions in the form of Laurent series. In addition, multiplier approach is used to obtain the conserved vectors and to analyzed conservations properties of the equation. We obtain four cases and the Conservation laws were verified for all the cases. \\
	\textbf{Keywords:} Kudryashov Sinelshchikov equation, Lie Symmetry Analysis, Painlevé analysis, Conservation laws, Multiplier Approach.
	\section{Introduction}
	Nonlinear partial differential equations (NLPDEs) plays a significant role in various complex processes that can be used in the modeling of physics, chemistry, biology, and engineering. They form the basis of the mathematical account of various nonlinear phenomena, including the shallow water flow, oscillation of a plasma, dynamic behaviour of optical pulses, and pressure wave propagation in gas liquid mixtures. These systems are frequently nonlinear and produce very dramatic wave patterns by solitary waves, shocks, breathers and rogue waves that remain a topic of interest both in nonlinear science and applied mathematics. This is why the study of NLPDEs and the creation of specific or approximate solutions are a significant component of the modern theoretical studies.
	\smallskip
	
	In the past few decades, a number of strong methods of analysis have been brought forth to obtain precise solutions or approximate solutions on nonlinear evolution equations (NLEEs). There are examples like classical techniques like Lie symmetry method~\cite{olver1993}, Painlevé analysis~\cite{weiss1983}, Hirota bilinear form~\cite{hirota1971}, Bäcklund and Darboux transformations~\cite{rogers1982,matveev1991}, and the inverse scattering transform~\cite{ablowitz1981} that have been found to be especially effective in the case of integrable systems. Simultaneously, expansion-based and sub-equation methods have been developed to address both integrable and non-integrable equations, such as the tanh-function method~\cite{malfliet1996}, sech–tanh expansion method~\cite{fan2000}, sine–cosine approach~\cite{wazwaz2005}, $(G'/G)$-expansion method~\cite{wang2008}, $F$-expansion method~\cite{zhang2011}, exp-function technique~\cite{he2006}, Jacobi elliptic function expansion~\cite{khater2006}, improved Riccati equation method~\cite{zayed2008}, first integral method~\cite{feng2002}, and the extended simplest equation approach~\cite{kudryashov2005}. 
	\smallskip
	
	The homogeneous balance technique~\cite{wang1996}, Cole–Hopf transformation~\cite{cole1951}, multiple-scale analysis~\cite{whitham1974}, variational iteration method (VIM)~\cite{he1999vim}, Adomian decomposition method (ADM)~\cite{adomian1994}, homotopy analysis method (HAM)~\cite{liao2003}, and homotopy perturbation method (HPM)~\cite{he1999hpm} and other schemes have also been applied successfully to obtain analytical or semi-analytical solutions.
	\smallskip
	
	In $19^{th}$ century, Norwegian mathematician Sophus Lie proposed the technique which helps in finding the exact solutions of differential equations called as Lie symmetry method. This method offers a structured way to determine continuous transformation groups that leave a differential equation unchanged. Symmetry obtained helps in reducing nonlinear PDE to one or more ordinary differential equations, which in many cases makes it possible to construct similarity forms or invariant solutions. In Painlev\'e analysis, we examine whether the given nonlinear differential equation is integrability by checking whether their general solution contain movable critical singularities. If the given PDE passes the Painlev\'e test, we say that the PDE is integrable and their solution form is given in Laurent series form. Along with these approaches,
	By the help of multiplier approach we derive the conservation laws for the PDE which play an important role in discovering physical invariants such as energy, momentum, and mass, and in validating analytical solutions.
	In addition to these classical methods, a broad variety of simple and complex analytic methods, such as a Hirota bilinear method, B\"acklund transformations and long-wave asymptotic expansions have been applied to find multi-soliton, breather, lump and rogue-wave solutions in most nonlinear evolution equations. These theories have enhanced the knowledge of higher-dimensional wave models of nonlinear interaction and stability properties.
	\smallskip
	
	Kudryashov and Sinelshchikov in 2010\cite{Kudryashov2010}, proposed a nonlinear partial differential equation which represents pressure waves propagating within a mixture of liquid and gas bubbles. The Kudryashov-Sinelshchikov(KS) Equation has become a key model to study the phenomenona of nonlinear acoustics, bubble dynamics, and dispersive wave propagation. The general form of (1+3)-diemnsional Kudryashov-Sinelshchikov(KS) Equation is given as follow
	\cite{Ali2021,Liu2021}
	\begin{equation}
		(u_t + \alpha u u_x + \gamma u_{xxx})_x + d u_{yy} + e u_{zz} = 0,
	\end{equation}
	where the function $u(x,y,z,t)$ represents a physical quantity such as density, pressure, or fluid velocity, and the arbitrary parameters $\alpha$, $\gamma$, $d$, and $e$ correspond to the nonlinear, dispersive, and diffusive parameters of the system.
	In Nuruzzaman \textit{et al.}\,\cite{nuruzzaman2023localized} reduced the (1+3)-diemnsional Kudryashov-Sinelshchikov(KS) Equation to $(1+2)$-dimensional form of the KS equation by substituting $z = x$, which is given as follow
	\begin{equation}
		(u_t + \alpha u u_x + \gamma u_{xxx})_x + d u_{yy} + e u_{xx} = 0.
	\end{equation}
	This reduced model retains the essential nonlinear and dispersive structure of the original equation and serves as a useful framework for the study of pressure wave in liquid and gas medium.
	\smallskip
	
	Nuruzzaman \textit{et al.}\,\cite{nuruzzaman2023localized} used Hirota bilinear method and constructed various families of solutions, including multisoliton, breather, lump, and mixed interaction patterns. Their numerical visualizations demonstrated a rich set of nonlinear dynamics and interactions between localized structures.
	\smallskip
	
	We can observe that obtaining the analytic solutions of given nonlinear PDEs as become fundamental. Hence, in the present study we apply three complementary analytical approaches -Lie symmetry analysis, the Painlev\'e test, and the conservation laws using multiplier approach which helps in conducting a detailed and systematic study about $(1+2)$-dimensional KS equation.
	\smallskip
	
	The structure of this paper is organized as follows: In Section~2, the Lie symmetry method is applied to determine the infinitesimal generators and the corresponding invariant solutions of the equation. Section~3 is devoted to performing the Painlevé analysis on the symmetry-reduced ordinary differential equation (ODE) to examine its integrability and to obtain analytical solutions in the Laurent series form, highlighting the singularity structure of the reduced equation. In Section~4, the conservation law approach is utilized to derive the conserved vectors and associated physical invariants of the system.
	
	\section{Lie Symmetry Analysis and Symmetry Reduction}
	In this section, we preform Lie Symmetry Analysis of  (1+2)-Dimensional Kudryashov-Sinelshchikov (KS) equation \cite{nuruzzaman2023localized}.
	The one-parameter Lie group of point transformations\cite{olver1993} under which Eq.(2) remains invariant, i.e.,
	\begin{eqnarray}
		t^{*}=t+\epsilon\tau(t,x,y,u)+\textit{O}(\epsilon^{2})\\ \nonumber 
		x^{*}=x+\epsilon\xi(t,x,y,u)+\textit{O}(\epsilon^{2})\\ \nonumber
			y^{*}=t+\epsilon\zeta(t,x,y,u)+\textit{O}(\epsilon^{2})\\ \nonumber 
		u^{*}=u+\epsilon\eta(t,x,y,u)+\textit{O}(\epsilon^{2})
	\end{eqnarray}
	where the group parameter is represented as $\epsilon$ and  infinitesimal of the transformations for independent and dependent variables are given as $\tau, \xi, \zeta, \eta$.
	\newline
	Infinitesimal generator of the given transformation is 
	\begin{equation}
		\textbf{X}=\tau(t,x,y,u)\dfrac{\partial}{\partial{t}}+\xi(t,x,y,u)\dfrac{\partial}{\partial{x}}+\zeta(t,x,y,u)\dfrac{\partial}{\partial{y}}+\eta(t,x,y,u)\dfrac{\partial}{\partial{u}}
	\end{equation}
	The infinitesimal generator $\textbf{X}$ must satisfy the following Lie invariance condition for PDE(2) which aids in determining the infinitesimals $\tau(t,x,y,u), \xi(t,x,y,u), \zeta(t,x,y,u), $  and $\eta(t,x,y,u)$.  
	\begin{equation}
		\textbf{X}^{(4)}(\triangle)\bigg|_{\triangle = 0}=0,
	\end{equation}
	where 
	\begin{equation}
		\triangle= du_{yy}+\alpha(u_{x})^2+eu_{xx}+\alpha uu_{xx}+\gamma u_{xxxx}+u_{tx}\nonumber
	\end{equation}
	and $\textbf{X}^{(4)}$ represents the fourth prolongation of the vector field $\textbf{X}$. The initial terms of this prolongation can be written in the form
	
	\begin{equation}
		\textbf{X}^{(4)}=\textbf{X}+\eta^{x}\dfrac{\partial}{\partial{u_{x}}}+\eta^{t}\dfrac{\partial}{\partial{u_{t}}}+\eta^{y}\dfrac{\partial}{\partial{u_{y}}}+\eta^{xx}\dfrac{\partial}{\partial{u_{xx}}}+\eta^{xy}\dfrac{\partial}{\partial{u_{xy}}}+\eta^{xt}\dfrac{\partial}{\partial{u_{xt}}}+\eta^{yy}\dfrac{\partial}{\partial{u_{yy}}}+\eta^{yt}\dfrac{\partial}{\partial{u_{yt}}}+\eta^{tt}\dfrac{\partial}{\partial{u_{tt}}}...
	\end{equation}
The fourth-order prolongation $\textbf{X}^{(4)}$ is applied to the given PDE(2) and we get,
	\begin{equation}
		\gamma \eta^{xxxx}+e\eta^{xx}+d\eta^{yy}+\eta^{xt}+2\alpha u_{x}\eta^{x}+\alpha u \eta^{xx}=0
	\end{equation}
	where $\eta^{x}
, \eta^{xx}, \eta^{xt}, \eta^{yy}, \eta^{xxxx}$ are the extended infinitesimals. We obtain the determining equations by expanding the Eq.(7)
\begin{align*}
	 &\zeta_{u}=0, \quad 
	 \xi_{u}=0, \quad 
	 \tau_{u}=0, \quad \tau_{x}=0, \quad 
	\eta_{uu}=0, \quad 
	 \\[6pt]
	& \gamma d \tau_{xy}=0, \quad 
	\gamma d \tau_{y}=0, \quad 
	\alpha d \tau_{y}=0, \quad 
	\eta_{xu}-d \tau_{yy}=0, \\[6pt]
	& 2d \tau_{y}+\zeta_{x}=0, \quad 
	\gamma \tau_{t}-3\gamma \xi_{x}=0, \quad 
	2\gamma d \tau_{yy}-3\gamma \xi_{xx}=0, \\[6pt]
	& d \tau_{t}+d \xi_{x}-2d \zeta_{y}=0, \quad 
	\alpha \tau_{t}+\alpha \eta_{u}-\alpha \xi_{x}=0, \\[6pt]
	& d \eta_{yy}+e \eta_{xx}+\eta_{tx}+\alpha u \eta_{xx}+\gamma \eta_{xxxx}=0, \\[6pt]
	& -2d \xi_{y}-2e \zeta_{x}-\zeta_{t}-2\alpha u \zeta_{x}-4\gamma \zeta_{xxx}=0, \\[6pt]
	& 2d \eta_{yu}-d \zeta_{yy}-e \zeta_{xx}-\zeta_{tx}-\alpha u \zeta_{xx}-\gamma \zeta_{xxxx}=0, \\[6pt]
	& \alpha \eta(t,x,y,u)+e \tau_{t}-e \xi_{x}-\xi_{t}+\alpha u \tau_{t}-\alpha u \xi_{x} \\
	& \qquad + 6\gamma \eta_{xxu}-4\gamma \xi_{xxx}=0, \\[6pt]
	& -d \xi_{yy}+2e \eta_{xu}-e \xi_{xx}+\eta_{tu}-\xi_{tx}+2\alpha u \eta_{xu}-\alpha u \xi_{xx} \\
	& \qquad + 4\gamma \eta_{xxxu}+2\alpha \eta_{x}-\gamma \xi_{xxxx}=0
\end{align*}
 By using mathematical software such as Maple or Mathematica we will solve the above system of determining equations.
 \newline
  We obtain the general symmetry vector of PDE(2), 
  \begin{equation}
  	\begin{split}
  		\text{dt} (\text{a}(t))+\text{dy} \left(\frac{2}{3} y \text{a}'(t)+\text{g}(t)\right)
  		\\+\text{dx} \left(-\frac{y^2 \text{a}''(t)}{6 d}+\frac{1}{3} x \text{a}'(t)-\frac{y \text{g}'(t)}{2 d}+\text{h}(t)\right)
  		\\+\text{du} \left(-\frac{y^2 \text{a}^{'''}(t)}{6 \alpha  d}+\frac{x \text{a}''(t)}{3 \alpha }+\frac{3 \text{h}'(t)-2 e \text{a}'(t)}{3 \alpha }-\frac{2}{3} u \text{a}'(t)-\frac{y \text{g}''(t)}{2 \alpha  d}\right)
  	\end{split}
  \end{equation}

	where $a(t)$, $g(t)$, and $h(t)$ denotes arbitrary functions of time. 
	The presence of these free functions leads to infinitely many admissible symmetry 
	generators, which implies that the (1+2)-dimensional Kudryashov--Sinelshchikov (KS) 
	equation possesses an infinite-dimensional Lie algebra.
	
	Each arbitrary function corresponds to an independent family of vector fields, and 
	the three principal generators associated with $a(t)$, $g(t)$, and $h(t)$ can be expressed as
\begin{align}
	X_{1} = & \left(
	-\frac{y^{2} a^{'''}(t)}{6 \alpha d}
	+ \frac{x a''(t)}{3 \alpha}
	- \frac{2 e a'(t)}{3 \alpha}
	- \frac{2}{3} u a'(t)
	\right)\frac{\partial}{\partial u} \nonumber \\[6pt]
	& + \left(
	-\frac{y^{2} a''(t)}{6 d}
	+ \frac{1}{3} x a'(t)
	\right)\frac{\partial}{\partial x}
	+ \frac{2}{3} y a'(t)\frac{\partial}{\partial y}
	+ a(t)\frac{\partial}{\partial t} \nonumber \\[10pt]
	X_2 = & -\frac{y g''(t)}{2 \alpha d}\frac{\partial}{\partial u}
	- \frac{y g'(t)}{2 d}\frac{\partial}{\partial x}
	+ g(t)\frac{\partial}{\partial y} \nonumber \\[10pt]
	X_3 = & \frac{h'(t)}{\alpha}\frac{\partial}{\partial u}
	+ h(t)\frac{\partial}{\partial x}
\end{align}

 By evaluating the commutative product between these vector fields we derive the optimal values for these functions.
 \newline
 Appling the commutative product $[X_{1},X_{2}]=X_{1}(X_{2})-X_{2}(X_{1})$ between these vectors and equating to zero leads to system of ODEs as follow:
 \begin{equation}
 	\begin{split}
 		a(t) h'-\frac{1}{3} a' h(t)=0,
 		\\
 		\frac{2 a' h'}{3 \alpha }-\frac{a^{\prime \prime } h(t)}{3 \alpha }+\frac{a(t) h^{\prime \prime }}{\alpha }=0,\\
 		y a(t) g'-\frac{2}{3} y a' g(t)=0,
 		\\
 		-\frac{a' g'}{6 d}+\frac{a^{\prime \prime } g(t)}{3 d}-\frac{a(t) g^{\prime \prime }}{2 d}=0,
 		\\
 		\frac{a^{\prime \prime } g'}{6 \alpha  d}-\frac{2 a' g^{\prime \prime }}{3 \alpha  d}+\frac{a^{\prime \prime \prime } g(t)}{3 \alpha  d}-\frac{a(t) g^{\prime \prime \prime }}{2 \alpha  d}=0
 	\end{split}
 \end{equation}
 With the help of Maple and manual calculations we solve the above system of nonlinear ordinary differential equations. We obtain
 \begin{equation}
 	\begin{split}
 		h(t)=c_{1}(a(t))^{1/3}\\
 		g(t)=c_{2}(a(t))^{2/3}
 	\end{split}
 \end{equation}
 We can see that $a(t)$ is the independent arbitrary function in time and $g(t)$ and $h(t)$ depends on $a(t)$.
\newline
Let us consider the value for $a(t)$ which will give the value for the $g(t)$ and $h(t)$. Therefore, we consider
\begin{equation}
	\begin{split}
		a(t)=(p_{1}t+p_{2})^{3}\\
		g(t)=(p_{1}t+p_{2})^{2}\\
	h(t)=(p_{1}t+p_{2})
	\end{split}
\end{equation}
where $p_{1}$ and $p_{2}$ are the arbitrary constants. Here, $a(t)$ is taken as a cubic polynomial in $t$ because it appears with derivatives up to the third order in symmetry $X_{1}$. Choosing the polynomial of third degree allows to maintain the consistency in analysis.
\newline
Now, by substituting the above value of functions into the three vector fields given by Eq.(9), we explore the three unknown Symmetries. They are
\begin{equation}
	\begin{aligned}
		X_1 = & \;\Bigg(
		-\frac{6 p_1^3 y^2}{6 \alpha d}
		- \frac{6 e p_1 (p_1 t + p_2)^2}{3 \alpha}
		+ \frac{6 p_1^2 x (p_1 t + p_2)}{3 \alpha}
		- 2 p_1 u (p_1 t + p_2)^2
		\Bigg)\frac{\partial}{\partial u} \\[2pt]
		& + \Bigg(
		-\frac{6 p_1^2 y^2 (p_1 t + p_2)}{6 d}
		+ p_1 x (p_1 t + p_2)^2
		\Bigg)\frac{\partial}{\partial x} \\[2pt]
		& + 2 p_1 y (p_1 t + p_2)^2 \frac{\partial}{\partial y} 
		+ (p_1 t + p_2)^3 \frac{\partial}{\partial t} \\[4pt]
		X_2 = & \; -\frac{2 p_1^2 y}{2 \alpha d} \frac{\partial}{\partial u} 
		- \frac{2 p_1 y (p_1 t + p_2)}{2 d} \frac{\partial}{\partial x} 
		+ (p_1 t + p_2)^2 \frac{\partial}{\partial y} \\[2pt]
		X_3 = & \;\frac{p_1}{\alpha} \frac{\partial}{\partial u} 
		+ (p_1 t + p_2) \frac{\partial}{\partial x}
	\end{aligned}
\end{equation}

	\subsection{Symmetry Reduction and Group-Invariant Solutions}
	To obtain the reduced forms of the governing equation, the characteristic system
	\begin{equation}
		\frac{du}{\eta} = \frac{dt}{\tau} = \frac{dx}{\xi} = \frac{dy}{\zeta},
	\end{equation}
	is integrated for each infinitesimal generator $X_i$ $(i=1,2,3)$. Solving these characteristic relations allows the original partial differential equation to be rewritten either as an ordinary differential equation or as a lower–order PDE, from which the corresponding group-invariant solutions can be determined.

	\subsubsection{Symmetry Reduction using $X_{1}$}
	Symmetry $X_{1}$ which is corresponding to the arbitrary function $a(t)$ is given as
	\begin{equation}
		\begin{aligned}
			X_1 = & \;\Bigg(
			-\frac{6 p_1^3 y^2}{6 \alpha d}
			- \frac{6 e p_1 (p_1 t + p_2)^2}{3 \alpha}
			+ \frac{6 p_1^2 x (p_1 t + p_2)}{3 \alpha}
			- 2 p_1 u (p_1 t + p_2)^2
			\Bigg)\frac{\partial}{\partial u} \\[2pt]
			& + \Bigg(
			-\frac{6 p_1^2 y^2 (p_1 t + p_2)}{6 d}
			+ p_1 x (p_1 t + p_2)^2
			\Bigg)\frac{\partial}{\partial x} \\[2pt]
			& + 2 p_1 y (p_1 t + p_2)^2 \frac{\partial}{\partial y} 
			+ (p_1 t + p_2)^3 \frac{\partial}{\partial t}
		\end{aligned}
	\end{equation}
	The charateristics equation of the symmetry $X_{1}$ is given as:
	\begin{equation}
		\begin{split}
			\frac{du}{-\frac{6 p_1^3 y^2}{6 \alpha d} 
				- \frac{6 e p_1 (p_1 t + p_2)^2}{3 \alpha} 
				+ \frac{6 p_1^2 x (p_1 t + p_2)}{3 \alpha} 
				- 2 p_1 u (p_1 t + p_2)^2} 
			= \frac{dt}{(p_1 t + p_2)^3} \\
			= \frac{dx}{-\frac{6 p_1^2 y^2 (p_1 t + p_2)}{6 d} + p_1 x (p_1 t + p_2)^2} 
			= \frac{dy}{2 p_1 y (p_1 t + p_2)^2}
		\end{split}
	\end{equation}
	We solve the above lagrangian equations which give rise to the similarity variable and simlarity solutions. Substituting the obtained similarity variable along with similarity solution into the given PDE(2), it will reduce to (1+1)-dimensional PDE which is given as
	\begin{equation}
		\begin{split}
			& \alpha^2 d f_{qq} f(l,q) 
			+ \alpha d^2 f_{ll} 
			+ \alpha d e p_2^2 f_{qq} 
			- 2 \alpha d l p_1 p_2^2 f_{lq}- 3 \alpha d p_1 p_2^2 f_{q} 
			- \alpha d p_1 p_2^2 q f_{qq}  \\
			&  
			+ \alpha^2 d f_{q}^2 
			+ \alpha \gamma d f_{qqqq} + \alpha l^2 p_1^2 p_2^4 f_{qq} 
			+ 2 d p_1^2 p_2^4 = 0
		\end{split}
	\end{equation}
	where $l$ and $q$ are the similarity variable given as
	\begin{equation}
		\begin{split}
				l=\frac{y}{\left(p_1 t+p_2\right){}^2}\\
					q=\frac{2 d p_1^3 t^3 x+6 d p_2 p_1^2 t^2 x+6 d p_2^2 p_1 t x+2 d p_2^3 x+p_1^3 t^2 y^2+2 p_2 p_1^2 t y^2}{2 d \left(p_1 t+p_2\right){}^4}
		\end{split}
	\end{equation}
	We can see that PDE (17) is fourth order nonlinear PDE which is not easily solvable.
	\newline
	Therefore, we proceed by finding the point symmetries of the PDE (17).
	After performing the series of steps as we done previously, we obtain three point symmetries of PDE(17) given as

	\begin{align}
		X_{11} &= \frac{p_{1} p_{2}^{2}}{\alpha}\,\frac{\partial}{\partial f}
		+ \frac{\partial}{\partial q}, \\[2mm]
		X_{12} &= -\frac{ l \, p_1^2 p_2^4}{\alpha d}\,\frac{\partial}{\partial f}
		- \frac{ l \, p_1 p_2^2}{d}\,\frac{\partial}{\partial q}
		+ \frac{\partial}{\partial l}, \\[2mm]
		X_{13} &= \Big(-\frac{3}{4 \alpha d} l^2 p_1^2 p_2^4 
		- \frac{1}{\alpha} e p_2^2 
		- f 
		+ \frac{3}{2 \alpha} p_1 p_2^2 q \Big) \frac{\partial}{\partial f} \notag \\
		&\quad + \Big(-\frac{3}{4 d} l^2 p_1 p_2^2 
		+ \frac{1}{2} q \Big) \frac{\partial}{\partial q} 
		+ l \frac{\partial}{\partial l}.
	\end{align}
	\subsubsection*{Symmetry Reduction using $X_{11}$}
	By considering the symmetry $X_{11}$, we again perform symmetry reduction by solving the characteristics equations which give rise to similarity solutions,
	\begin{equation}
		f(l,q)=\mathcal{F}_1(l)+\frac{p_1 p_2^2 q}{\alpha }
	\end{equation}
	where $l$ and $q$ are the similarity variables given in Eq.(18).
	\newline
	Substituting the Eq.(22) into the symmetry-reduced PDE (17), we obtain a second order ODE which is easily solvable
	\begin{equation}
		\alpha  d^2 \mathcal{F}_1{}^{\prime \prime }=0
	\end{equation}
	solution of above ODE is given as
	\begin{equation}
		\mathcal{F}_1=K_{1}+qK_{2}
	\end{equation}
	where $K_{1}$ and $K_{2}$ are arbitrary constants. 
	
	\subsubsection*{Symmetry Reduction using $X_{12}$}
	By considering the symmetry $X_{12}$, we again perform symmetry reduction by solving the characteristics equations which give rise to similarity variable,
	\begin{equation}
		v = \frac{2 d q+l^2 p_1 p_2^2}{2 d}
	\end{equation}
	and similarity solutions
	\begin{equation}
			f(l,q)=\frac{2 \alpha  d \psi(v)-l^2 p_1^2 p_2^4}{2 \alpha  d}
	\end{equation}
	where $l$ and $q$ are the similarity variables given in Eq.(18).
	\newline
	Substituting the Eq.(26) along with Eq.(25) into the symmetry-reduced PDE (17), we obtain a fourth order nonlinear ODE which is not easily solvable.
	\begin{equation}
		\alpha ^2 d \psi '^2+\alpha  \gamma  d \psi ^{\prime \prime \prime \prime }+\alpha  d e p_2^2 \psi ^{\prime \prime }-2 \alpha  d p_1 p_2^2 \psi '-\alpha  d p_1 p_2^2 v \psi ^{\prime \prime }+d p_1^2 p_2^4+\alpha ^2 d \psi (v) \psi ^{\prime \prime }=0
	\end{equation}
	We proceed by obtaining the solution of above nonlinear ODE (27) by Painlev\'e analysis.

	\subsubsection*{Symmetry Reduction using $X_{13}$}
	By considering the symmetry $X_{13}$, we again perform symmetry reduction by solving the characteristics equations which give rise to similarity variable,
	\begin{equation}
		w = \frac{2 d q+l^2 p_1 p_2^2}{2 d \sqrt{l}}
	\end{equation}
	and similarity solutions
	\begin{equation}
		f(l,q)=\frac{\alpha  s(w)-e l p_2^2+l p_1 p_2^2 q}{\alpha  l}
	\end{equation}
	where $l$ and $q$ are the similarity variables given in Eq.(18).
	\newline
	Substituting the Eq.(29) along with Eq.(28) into the symmetry-reduced PDE (17), we obtain a fourth order nonlinear ODE which is not easily solvable.
	\begin{equation}
		\alpha  d^2 w^2 s^{\prime \prime }+7 \alpha  d^2 w s'+8 \alpha  d^2 s(w)+4 \alpha ^2 d s'^2+4 \alpha  \gamma  d s^{\prime \prime \prime \prime }+4 \alpha ^2 d s(w) s^{\prime \prime }=0
	\end{equation}
	We proceed by obtaining the solution of above nonlinear ODE (30) by Painlev\'e analysis.
	
		\subsubsection{Symmetry Reduction using $X_{2}$}
	Symmetry $X_{2}$ which is corresponding to the arbitrary function $g(t)$ is given as
	\begin{equation}
		\begin{aligned}
				X_2 = & \; -\frac{2 p_1^2 y}{2 \alpha d} \frac{\partial}{\partial u} 
			- \frac{2 p_1 y (p_1 t + p_2)}{2 d} \frac{\partial}{\partial x} 
			+ (p_1 t + p_2)^2 \frac{\partial}{\partial y}
		\end{aligned}
	\end{equation}
	The charateristics equation of the symmetry $X_{2}$ is given as:
	\begin{equation}
		\begin{split}
			\frac{du}{-\frac{2 p_1^2 y}{2 \alpha d}} 
			= \frac{dt}{0} 
			= \frac{dx}{- \frac{2 p_1 y (p_1 t + p_2)}{2 d}}
			= \frac{dy}{(p_1 t + p_2)^2}
		\end{split}
	\end{equation}
	We solve the above lagrangian equations which give rise to the similarity variable and simlarity solutions. Substituting the obtained similarity variable along with similarity solution into the given PDE(2), it will reduce to (1+1)-dimensional PDE which is given as
	\begin{equation}
		\begin{split}
			&4 x^2 \phi _{\text{, }\omega \omega } \left(\alpha  \left(y^2-2 \omega \right)^2 \phi (t,\omega )+8 d x^2 \omega +e \left(y^2-2 \omega \right)^2\right)+64 d x^4 \phi _{\text{, }\omega }\\
			&
			+\left(y^2-2 \omega \right) \left(\gamma  \left(y^2-2 \omega \right)^3 \phi _{\text{, }\omega \omega \omega \omega }-8 x^3 \phi _{\text{, }t\omega }\right)+4 \alpha  x^2 \left(y^2-2 \omega \right)^2 \phi _{\text{, }\omega }{}^2=0
		\end{split}
	\end{equation}
	where $t$ and $\omega$ are the similarity variables  given as
	\begin{equation}
		\begin{split}
			\omega=\frac{2 d p_1 t x+2 d p_2 x+p_1 y^2}{2 p_1}
		\end{split}
	\end{equation}
	and similarity solution
	\begin{equation}
	u=	\frac{\alpha  p_1 t \phi \left(t,\omega \right)+\alpha  p_2 \phi\left(t,\omega\right)+p_1 x}{\alpha  \left(p_1 t+p_2\right)}
	\end{equation}
	We can see that PDE (33) is fourth order nonlinear PDE which is not easily solvable.
	\newline
	Therefore, we proceed by finding the point symmetries of the PDE (33).
	After performing the series of steps as we done previously, we obtain one point symmetries of PDE(33) given as
	
	\begin{align}
		X_{21} &=\frac{\partial}{\partial t}.
	\end{align}
	\subsubsection*{Symmetry Reduction using $X_{21}$}
	By considering the symmetry $X_{21}$, we again perform symmetry reduction by solving the characteristics equations which give rise to similarity solutions,
	\begin{equation}
		\phi (t,\omega )=\mu (\omega )
	\end{equation}
	where $\omega$ is the similarity variables given in Eq.(34).
	\newline
	Substituting the Eq.(37) into the symmetry-reduced PDE (33), we obtain a fourth order nonlinear ODE which is not easily solvable.
	\begin{equation}
		\begin{split}
			&4 x^2 \mu ^{\prime \prime } \left(8 d x^2 \omega +e \left(y^2-2 \omega \right)^2+\alpha  \mu (\omega ) \left(y^2-2 \omega \right)^2\right)+64 d x^4 \mu '\\
			&+4 \alpha  x^2 \mu '^2 \left(y^2-2 \omega \right)^2+\gamma  \mu ^{\prime \prime \prime \prime } \left(y^2-2 \omega \right)^4=0
		\end{split}
	\end{equation}
		We proceed by obtaining the solution of above nonlinear ODE (38) by Painlev\'e analysis.
	\subsubsection{Symmetry Reduction using $X_{3}$}
	Symmetry $X_{3}$ which is corresponding to the arbitrary function $h(t)$ is given as
	\begin{equation*}
		\begin{aligned}
			X_3 = & \;\frac{p_1}{\alpha} \frac{\partial}{\partial u} 
			+ (p_1 t + p_2) \frac{\partial}{\partial x}
		\end{aligned}
	\end{equation*}
	The charateristics equation of the symmetry $X_{3}$ is given as:
	\begin{equation}
		\frac{du}{\frac{p_1}{\alpha}}=\frac{dt}{0}=\frac{dx}{(p_1 t + p_2)}=\frac{dy}{0}
	\end{equation}
	We solve the above lagrangian equations which give rise to the similarity solution,
	\begin{equation}
		u = \mathcal{F}_2(t,y)+\frac{p_1 x}{\alpha  \left(p_1 t+p_2\right)}
	\end{equation}
	Substituting (40) into given PDE(2) we obtain,
	\begin{equation}
		d \mathcal{F}_{2,yy}=0
	\end{equation}
	We can see that, the above PDE is easily solvable. The solution of (41) is given as,
	\begin{equation}
		\mathcal{F}_{2}=y\phi_{1}(t)+\phi_{2}(t)
	\end{equation}
	where $\phi_{1}(t)$ and $\phi_{2}(t)$ are arbitrary functions in time.
	
	\section{Painlev\'e Analysis}
	In the previous section, we obtain fourth order nonlinear ordinary differential equations which we will solve in this section with the aids of Painlev\'e Analysis given by ARS Algorithm \cite{Lakshmanan1993,Ramani1989}. ARS algorithm have three major steps. Firstly, we will find the leading-order behaviour then we will compute the resonances and lastly we will evaluate for the existence of sufficient number of arbitrary constants in the Laurent series expansion. If an equation passes these main steps in the ARS algorithm, it is termed to be painlev\'e integrable.
	\subsection{Painlev\'e Analysis of ODE (27)}
First step is to assume that in a sufficiently small region around a movable singular point \( v = v_0 \) in the complex plane, the dominant behavior of the solution can be approximated by
\begin{equation}
	\psi(v) \approx a_0 (v - v_0)^n,
\end{equation}
where \( a_0 \neq 0 \) is a constant and \( n \in \mathcal{Z}\) is the leading-order exponent.
\newline
Substituting the Eq.(43) into the Symmetry-Reduced ODE(27), we obtain
\begin{equation}
	\begin{split}
		\psi(v) =&\; 2 a_0^{2} \alpha^{2} d n^{2} (v-v_0)^{2n-2}
		- a_0^{2} \alpha^{2} d n (v-v_0)^{2n-2}
		+ a_0 \alpha d e n^{2} p_2^{2} (v-v_0)^{n-2} \\
		&- a_0 \alpha d e n p_2^{2} (v-v_0)^{n-2}
		- a_0 \alpha d n^{2} p_1 p_2^{2} v (v-v_0)^{n-2}
		+ a_0 \alpha d n p_1 p_2^{2} v (v-v_0)^{n-2} \\
		&- 2 a_0 \alpha d n p_1 p_2^{2} (v-v_0)^{n-1}
		+ a_0 \alpha \gamma d n^{4} (v-v_0)^{n-4}
		- 6 a_0 \alpha \gamma d n^{3} (v-v_0)^{n-4} \\
		&+ 11 a_0 \alpha \gamma d n^{2} (v-v_0)^{n-4}
		- 6 a_0 \alpha \gamma d n (v-v_0)^{n-4}
		+ d p_1^{2} p_2^{4}.
	\end{split}
\end{equation}

By balnacing the highest order derivative and nonlinear term in the above equation we obtain the integral value of $n=-2$ and leading order coefficient $a_{0}=-\frac{12 \gamma }{\alpha }$.
\newline
To determine the resonances, which indicate the exponents at which arbitrary constants appear in the series solution, we substitute the perturbed form  
\begin{equation}
	\psi(v)  = a_{0}(v - v_{0})^{n} + m(v - v_{0})^{n+r}
\end{equation}
into the reduced ODE. By isolating the coefficients of $m$ and retaining the leading singular contributions, we obtain a biquadratic polynomial in $r$:
\begin{equation*}
	Q(r) = -120 \alpha  \gamma  d+\alpha  \gamma  d r^4-14 \alpha  \gamma  d r^3+59 \alpha  \gamma  d r^2-46 \alpha  \gamma  d r .
\end{equation*}
Setting $Q(r)=0$ gives the resonances
\begin{equation*}
	r = -1, \; 4,\; 5, \; 6 .
\end{equation*}
Here, $r=-1$ is the generic resonance corresponding to the arbitrariness of the singularity position $v_0$, while the non-generic resonances $r=4,\; 5$ and $6$ must be checked for consistency within the series expansion.
\newline
Therefore, we substitute the following Laurent expansion into the symmetry-reduced ODE:
\begin{equation}
	\begin{split}
		\psi(v) = &\frac{a_0}{(v-v_{0})^2} + \frac{a_1}{v-v_{0}} + a_{2} + a_3 (v-v_{0}) + a_4 (v-v_{0})^2 \\
		&+ a_5 (v-v_{0})^3+a_6 (v-v_{0})^4+a_7 (v-v_{0})^5+a_8 (v-v_{0})^6+a_9 (v-v_{0})^7+ \cdots,
	\end{split}
\end{equation}
\newline
By collecting the coefficients of $\tfrac{1}{(v - v_{0})^n}$ for $n=0,1,2,3,\dots$ and setting them equal to zero, we obtain the following relations:
\begin{equation}
	\begin{split}
		a_0 &= -\frac{12 \gamma }{\alpha }, \\
		a_1 &= 0, \\
		a_2 &= \frac{b p_1 p_2^2-e p_2^2}{\alpha }, \\
		a_3 &= \frac{p_1 p_2^2}{\alpha }. \\
	\end{split}
\end{equation}
\newline
In a similar way, the higher-order coefficients $a_7, a_8, a_9, \dots$ can be determined. It is observed that the coefficients of the series depend on $a_4, a_5,$ and $a_6$, which correspond to the arbitrary constants arising at the resonance values $4, 5,$ and $6$.  
\newline
This confirms that the right Painlev\'e series, together with the previously obtained leading-order behavior, represents the general solution of the considered equation.
\subsection{Painlev\'e Analysis of ODE (30)}
\noindent
We now apply the Painlev\'e analysis to the symmetry-reduced ODE Eq.(30) 
following the same procedure as before. 
First, we assume that near a movable singularity $w=w_0$, the dominant behavior of the solution can be expressed as
\begin{equation}
	s(w) \approx a_0 (w-w_0)^n,
\end{equation}
where \( a_0 \neq 0 \) is a constant and \( n \in \mathcal{Z}\) is the leading-order exponent.
\newline
Substituting the Eq.(48) into the Symmetry-Reduced ODE(30), we obtain
\begin{equation}
	\begin{split}
	s(w)=	&\alpha a_0 d^2 n^2 w^2 (w - w_0)^{\,n-2} 
		- \alpha a_0 d^2 n w^2 (w - w_0)^{\,n-2} 
		+ 7 \alpha a_0 d^2 n w (w - w_0)^{\,n-1} \\
		&+ 8 \alpha a_0 d^2 (w - w_0)^n
		+ 4 \alpha a_0 \gamma d n^4 (w - w_0)^{\,n-4} 
		- 24 \alpha a_0 \gamma d n^3 (w - w_0)^{\,n-4} \\
		&+ 8 \alpha^2 a_0^2 d n^2 (w - w_0)^{\,2n-2} 
		+ 44 \alpha a_0 \gamma d n^2 (w - w_0)^{\,n-4} 
		- 4 \alpha^2 a_0^2 d n (w - w_0)^{\,2n-2} \\
		&- 24 \alpha a_0 \gamma d n (w - w_0)^{\,n-4}
	\end{split}
\end{equation}
By balancing the highest derivative with the most singular nonlinear terms, we determine
\begin{equation*}
	n = -2, \quad
	a_0 = -\frac{12 \gamma }{\alpha }.
\end{equation*}
\newline
To compute the resonances, we consider a perturbed solution
\begin{equation}
	s(w) = a_0 (w-w_0)^n + m (w-w_0)^{n+r},
\end{equation}
and substitute it into the ODE(30). By collecting terms linear in $m$ and keeping the leading-order singular contributions, we obtain the resonance polynomial:
\begin{equation*}
	Q(r) = -480 \alpha  \gamma  d+4 \alpha  \gamma  d r^4-56 \alpha  \gamma  d r^3+236 \alpha  \gamma  d r^2-184 \alpha  \gamma  d r
\end{equation*}
Solving $Q(r)=0$ gives the resonance values:
\begin{equation*}
	r =  -1, 4, 5, 6
\end{equation*}
Here, $r=-1$ is the generic resonance corresponding to the arbitrariness of the singularity location $w_0$, and the positive integer resonances  $r=4,\; 5$ and $6$  correspond to positions where arbitrary constants may appear, which must be checked for consistency.
\newline
Based on the leading-order analysis, we construct the Laurent expansion:
\begin{equation}
	\begin{split}
		s(w) = & \frac{a_0}{(w-w_0)^2} + \frac{a_1}{(w-w_0)^{1}} + a_2 + a_3 (w-w_0) + a_4 (w-w_0)^2 \\
		& + a_5 (w-w_0)^3 + a_6 (w-w_0)^4 + a_7 (w-w_0)^5 + a_8 (w-w_0)^6 + \cdots,
	\end{split}
\end{equation}
\newline
Substituting the series \(s(w)\) into the symmetry reduced ODE and equating the coefficients of each power of \((w-w_0)\) to zero yields recurrence relations for the coefficients \(a_j\). Solving these relations gives the following values for the first few coefficients:

\begin{equation}
	\begin{aligned}
		a_0 &= -\frac{12 \gamma}{\alpha}, \\
		a_1 &= 0, \\
		a_2 &= -\frac{b^2 d}{4 \alpha}, \\
		a_3 &= \frac{b d}{4 \alpha},
	\end{aligned}
\end{equation}
while the coefficients \(a_4, a_5, a_6\) remain arbitrary.  
\newline
Thus, the Laurent expansion, together with the leading-order behavior, represents the general solution of the ODE and confirms the Painlev\'e property for this equation.
\subsection{Painlev\'e Analysis of ODE (38)}
\noindent
We now apply the Painlev\'e analysis to the symmetry-reduced ODE Eq.(38) 
following the same procedure as before. 
First, we assume that near a movable singularity $\omega=\omega_0$, the dominant behavior of the solution can be expressed as
\begin{equation}
	\mu(\omega) \approx a_0 (\omega-\omega_0)^n,
\end{equation}
where \( a_0 \neq 0 \) is a constant and \( n \in \mathcal{Z}\) is the leading-order exponent.
\newline
Substituting the Eq.(53) into the Symmetry-Reduced ODE(38), we obtain
\begin{equation}
	\begin{split}
		\mu(\omega) =&\, 16 n^4 \gamma \omega^4 a_0 p^{n-4} 
		- 96 n^3 \gamma \omega^4 a_0 p^{n-4} 
		+ 176 n^2 \gamma \omega^4 a_0 p^{n-4} 
		- 96 n \gamma \omega^4 a_0 p^{n-4} \\[4pt]
		& - 32 n^4 y^2 \gamma \omega^3 a_0 p^{n-4} 
		+ 192 n^3 y^2 \gamma \omega^3 a_0 p^{n-4} 
		- 352 n^2 y^2 \gamma \omega^3 a_0 p^{n-4} 
		+ 192 n y^2 \gamma \omega^3 a_0 p^{n-4} \\[4pt]
		& + 24 n^4 y^4 \gamma \omega^2 a_0 p^{n-4} 
		- 144 n^3 y^4 \gamma \omega^2 a_0 p^{n-4} 
		+ 264 n^2 y^4 \gamma \omega^2 a_0 p^{n-4} 
		- 144 n y^4 \gamma \omega^2 a_0 p^{n-4} \\[4pt]
		& + n^4 y^8 \gamma a_0 p^{n-4} 
		- 6 n^3 y^8 \gamma a_0 p^{n-4} 
		+ 11 n^2 y^8 \gamma a_0 p^{n-4} 
		- 6 n y^8 \gamma a_0 p^{n-4} \\[4pt]
		& - 8 n^4 y^6 \gamma \omega a_0 p^{n-4} 
		+ 48 n^3 y^6 \gamma \omega a_0 p^{n-4} 
		- 88 n^2 y^6 \gamma \omega a_0 p^{n-4} 
		+ 48 n y^6 \gamma \omega a_0 p^{n-4} \\[4pt]
		& + 4 e n^2 x^2 y^4 a_0 p^{n-2} 
		- 4 e n x^2 y^4 a_0 p^{n-2} 
		+ 16 e n^2 x^2 \omega^2 a_0 p^{n-2} 
		- 16 e n x^2 \omega^2 a_0 p^{n-2} \\[4pt]
		& + 32 d n^2 x^4 \omega a_0 p^{n-2} 
		- 32 d n x^4 \omega a_0 p^{n-2} 
		- 16 e n^2 x^2 y^2 \omega a_0 p^{n-2} 
		+ 16 e n x^2 y^2 \omega a_0 p^{n-2} \\[4pt]
		& + 64 d n x^4 a_0 p^{n-1} \\[4pt]
		& + 32 n^2 x^2 \alpha \omega^2 a_0^2 p^{2n-2} 
		- 16 n x^2 \alpha \omega^2 a_0^2 p^{2n-2} 
		+ 8 n^2 x^2 y^4 \alpha a_0^2 p^{2n-2} 
		- 4 n x^2 y^4 \alpha a_0^2 p^{2n-2} \\[4pt]
		& - 32 n^2 x^2 y^2 \alpha \omega a_0^2 p^{2n-2} 
		+ 16 n x^2 y^2 \alpha \omega a_0^2 p^{2n-2}.
	\end{split}
\end{equation}

By balancing the highest derivative with the most singular nonlinear terms, we determine
\begin{equation*}
	n = -2, \quad
	a_0 = -\frac{3 \gamma  \left(2 b-y^2\right)^2}{\alpha  x^2}.
\end{equation*}
\newline
To compute the resonances, we consider a perturbed solution
\begin{equation}
	\mu(\omega) = a_0 (\omega-\omega_0)^n + m (\omega-\omega_0)^{n+r},
\end{equation}
and substitute it into the ODE(38). By collecting terms linear in $m$ and keeping the leading-order singular contributions, we obtain the resonance polynomial:
\begin{equation*}
	\begin{split}
		Q(r) ={}& -1920 b^4 \gamma 
		+ 16 b^4 \gamma r^4 
		- 224 b^4 \gamma r^3 
		+ 944 b^4 \gamma r^2 
		- 736 b^4 \gamma r \\[4pt]
		& - 32 b^3 \gamma r^4 y^2 
		+ 448 b^3 \gamma r^3 y^2 
		- 1888 b^3 \gamma r^2 y^2 
		+ 1472 b^3 \gamma r y^2 
		+ 3840 b^3 \gamma y^2 \\[4pt]
		& + 24 b^2 \gamma r^4 y^4 
		- 336 b^2 \gamma r^3 y^4 
		+ 1416 b^2 \gamma r^2 y^4 
		- 1104 b^2 \gamma r y^4 
		- 2880 b^2 \gamma y^4 \\[4pt]
		& - 8 b \gamma r^4 y^6 
		+ 112 b \gamma r^3 y^6 
		- 472 b \gamma r^2 y^6 
		+ 368 b \gamma r y^6 
		+ 960 b \gamma y^6 \\[4pt]
		& + \gamma r^4 y^8 
		- 14 \gamma r^3 y^8 
		+ 59 \gamma r^2 y^8 
		- 46 \gamma r y^8 
		- 120 \gamma y^8.
	\end{split}
\end{equation*}
Solving $Q(r)=0$ gives the resonance values:
\begin{equation*}
	r =  -1, 4, 5, 6
\end{equation*}
Here, $r=-1$ is the generic resonance corresponding to the arbitrariness of the singularity location $\omega_0$, and the positive integer resonances  $r=4,\; 5$ and $6$  correspond to positions where arbitrary constants may appear, which must be checked for consistency.
\newline
Based on the leading-order analysis, we construct the Laurent expansion:
\begin{equation}
	\begin{split}
		\mu(\omega) ={}& \frac{a_0}{(\omega - \omega_0)^2} 
		+ \frac{a_1}{(\omega - \omega_0)^{1}} 
		+ a_2 
		+ a_3 (\omega - \omega_0) 
		+ a_4 (\omega - \omega_0)^2 \\[4pt]
		& + a_5 (\omega - \omega_0)^3 
		+ a_6 (\omega - \omega_0)^4 
		+ a_7 (\omega - \omega_0)^5 
		+ a_8 (\omega - \omega_0)^6 + \cdots.
	\end{split}
\end{equation}
\newline
After substituting the Laurent series into the reduced ordinary differential equation 
and equating the coefficients of each power of $(\omega - \omega_0)$ to zero, 
we obtain recurrence relations for the coefficients $a_j$. 
\newline
By following the same procedure, we find that the arbitrary constants 
corresponding to $a_4$, $a_5$, and $a_6$ fail to satisfy the compatibility conditions. 
Hence, the resonance conditions are not consistent, and the equation fails the Painlev\'e test and therefore we cannot obtain a solution in the form of a Laurent series. 
This shows that the given equation does not possess the Painlev\'e property.
\newline
It is observe that, if we substitute $d=0$ and $\alpha\ne0$, the compatibility conditions at resonances is satisfied. Therefore, it passes Painlev\'e test and we can obtain a solution in the form of a Laurent series. However, after substituting $d=0$, the governing equation reduces to (1+1)-dimensional PDE with $u=u(x,t)$. 
\newline
Similarily, when $\gamma=0$ and $\alpha\ne0$, the compatibility conditions at resonances is satisfied. Therefore, it passes Painlev\'e test and we can obtain a solution in the form of a Laurent series. In this case, the highest-order term in the governing equation vanishes.
\section{Conservation Laws using the multiplier approach}
The multiplier method is a systematic and widely used approach for constructing conservation laws 
of nonlinear partial differential equations (PDEs)\cite{AncoBluman2002a,AncoBluman2002b}. It is based on the principle that 
for a given system of PDEs, a linear combination of the equations, multiplied by appropriately chosen functions 
(called multipliers), can be expressed as a divergence. This allows one to identify conserved densities 
and associated fluxes.

Consider a partial differential equation (PDE) involving three independent variables 
$x, y, t$ and one dependent variable $u(x, y, t)$, written in the general form
\begin{equation}
	\Delta(x, y, t, u, u_{(1)}, u_{(2)}, \dots, u_{(n)}) = 0,
\end{equation}
where $u_{(k)}$ denotes all derivatives of $u$ up to order $k$.

A function $\Lambda(x, y, t, u, u_{(1)}, \dots)$ is called a \emph{multiplier} if the product
\begin{equation}
	\Lambda \, \Delta
\end{equation}
can be expressed as a total divergence, that is,
\begin{equation}
	\Lambda \, \Delta = D_t(\Phi^t) + D_x(\Phi^x) + D_y(\Phi^y),
\end{equation}
where $D_t$, $D_x$, and $D_y$ denote the total derivative operators with respect to 
$t$, $x$, and $y$, respectively. The functions $\Phi^t$, $\Phi^x$, and $\Phi^y$ are the 
components of the conserved flux vector.

The relation above defines a \emph{conservation law} for the PDE, which can be written in divergence form as
\begin{equation}
	D_t(\Phi^t) + D_x(\Phi^x) + D_y(\Phi^y) = 0,
\end{equation}
holding on all solutions $u(x, y, t)$ of $\Delta = 0$.
The components $(\Phi^t, \Phi^x, \Phi^y)$ represent the conserved densities and fluxes, 
while $\Lambda$ serves as the corresponding conservation law multiplier.
\newline
We will apply Multiplier approach\cite{Naz2012} to derive the conservation laws for (1+2)-Dimensional Kudryashov-Sinelshchikov (KS) equation.
\newline
First step is to obtain the determining equation for the multiplier $\Lambda(x,y,t,u)$ which is obtained by taking the variational derivative which is given as
\begin{equation}
	\frac{\delta}{\delta u}[\Lambda(du_{yy}+\alpha(u_{x})^2+eu_{xx}+\alpha uu_{xx}+\gamma u_{xxxx}+u_{tx})]=0
\end{equation}
where $\frac{\delta}{\delta u}$ is the Euler operator given as
\begin{equation}
	\frac{\delta}{\delta u} = \frac{\partial}{\partial u}
	- D_x \left( \frac{\partial}{\partial u_x} \right)
	+ D_x^2 \left( \frac{\partial}{\partial u_{xx}} \right)
	+ D_y^2 \left( \frac{\partial}{\partial u_{yy}} \right)
	+ D_x D_t \left( \frac{\partial}{\partial u_{xt}} \right)
	- \cdots
\end{equation}
Here, \( D_t \), \( D_x \), and \( D_y \) denote the total derivative operators with respect to
\( t \), \( x \), and \( y \), respectively, which are given by
\begin{align}
	D_x &= \frac{\partial}{\partial x}
	+ u_x \frac{\partial}{\partial u}
	+ u_{xy} \frac{\partial}{\partial u_y}
	+ u_{xt} \frac{\partial}{\partial u_t}
	+ u_{xx} \frac{\partial}{\partial u_x}
	+ \cdots, \\[6pt]
	D_y &= \frac{\partial}{\partial y}
	+ u_y \frac{\partial}{\partial u}
	+ u_{xy} \frac{\partial}{\partial u_x}
	+ u_{yt} \frac{\partial}{\partial u_t}
	+ u_{yy} \frac{\partial}{\partial u_y}
	+ \cdots, \\[6pt]
	D_t &= \frac{\partial}{\partial t}
	+ u_t \frac{\partial}{\partial u}
	+ u_{tx} \frac{\partial}{\partial u_x}
	+ u_{ty} \frac{\partial}{\partial u_y}
	+u_{tt} \frac{\partial}{\partial u_t}
	+ \cdots.
\end{align}
Expanding the Eq.(61) with the use of Eq.(63) to Eq.(65) then simplifying takes the following form:
\begin{equation}
	\begin{split}
		&\alpha \Lambda_{uu} u_x^2 u 
		+ 2 \alpha u_x \Lambda_{xu} u
		+ 2 \alpha \Lambda_{u} u_{xx} u
		+ \alpha \Lambda_{xx} u
		+ 2 d \Lambda_{u} u_{yy}
		+ d \Lambda_{uu} u_y^2
		+ 2 d u_y \Lambda_{yu}
		+ d \Lambda_{yy} \\[6pt]
		&+ e \Lambda_{uu} u_x^2
		+ 2 e u_x \Lambda_{xu}
		+ 2 e \Lambda_{u} u_{xx}
		+ e \Lambda_{xx}
		+ u_t \Lambda_{uu} u_x
		+ u_x \Lambda_{tu}
		+ 2 \Lambda_{u} u_{tx}
		+ u_t \Lambda_{xu}
		+ \Lambda_{tx} \\[6pt]
		&+ \alpha \Lambda_{u} u_x^2
		+ \gamma \Lambda_{uuuu} u_x^4
		+ 4 \gamma u_x^3 \Lambda_{xuuu}
		+ 6 \gamma \Lambda_{uuu} u_{xx} u_x^2
		+ 6 \gamma u_x^2 \Lambda_{xxuu}
		+ 4 \gamma \Lambda_{uu} u_{xxx} u_x \\[6pt]
		&+ 12 \gamma u_{xx} u_x \Lambda_{xuu}
		+ 4 \gamma u_x \Lambda_{xxxu}
		+ 2 \gamma \Lambda_{u} u_{xxxx}
		+ 3 \gamma \Lambda_{uu} u_{xx}^2 \\[6pt]
		&+ 4 \gamma u_{xxx} \Lambda_{xu}
		+ 6 \gamma u_{xx} \Lambda_{xxu}
		+ \gamma \Lambda_{xxxx} = 0
	\end{split}
\end{equation}
From the above Eq.(66), set of determining equations are obtain for finding the multipliers:
\begin{equation}
	\begin{split}
		&2 d \Lambda_{u} = 0, \quad 2 \gamma \Lambda_{u} = 0, \quad 2 \Lambda_{u} = 0, \quad d \Lambda_{uu} = 0, \quad \Lambda_{uu} = 0, \\[1mm]
		&3 \gamma \Lambda_{uu} = 0, \quad 4 \gamma \Lambda_{uu} = 0, \quad 6 \gamma \Lambda_{uuu} = 0, \quad \gamma \Lambda_{uuuu} = 0, \quad 2 d \Lambda_{yu} = 0, \\[1mm]
		&4 \gamma \Lambda_{xu} = 0, \quad \Lambda_{xu} = 0, \quad 12 \gamma \Lambda_{xuu} = 0, \quad 4 \gamma \Lambda_{xuuu} = 0, \\[1mm]
		&\alpha e \Lambda_{uu}^2 u + \alpha \Lambda_{u} + 6 \gamma \Lambda_{xxuu} = 0, \quad
		2 \alpha \Lambda_{u} u + 2 e \Lambda_{u} + 6 \gamma \Lambda_{xxu} = 0, \\[1mm]
		&2 \alpha \Lambda_{xu} u + 2 e \Lambda_{xu} + \Lambda_{tu} + 4 \gamma \Lambda_{xxxu} = 0, \quad
		\alpha \Lambda_{xx} u + d \Lambda_{yy} + e \Lambda_{xx} + \Lambda_{tx} + \gamma \Lambda_{xxxx} = 0.
	\end{split}
\end{equation}
And solving the above system we get
\begin{equation}
	\Lambda = -\frac{y^2 c_1'(t)}{2 d} + x\, c_1(t) + c_3(t) + x y K_{1}+ y\, c_2(t)
\end{equation}
where $K_{1}$ is arbitrary parameter and $c_{1}(t), c_{2}(t)$ and $c_{3}(t)$ are the arbitrary functions in $t$.
The required conserved quantities are then found with the aid of divergence identity given as:
\begin{equation}
	D_{t}\Phi^{t}+D_{x}\Phi^{x}+	D_{y}\Phi^{y}=\Lambda(du_{yy}+\alpha(u_{x})^2+eu_{xx}+\alpha uu_{xx}+\gamma u_{xxxx}+u_{tx})
\end{equation}
where $\Phi^{t}$ is the conserved density and $\Phi^{x}$ and $\Phi^{y}$ are the spatial flux in $x$-direction and $y$-directtion respectively.
\newline
\textbf{Case 1: Conservation law corresponding to the multiplier $\Lambda_{1} = xy$} 
\newline 
For the first multiplier associated with $K_{1}$, we consider $\Lambda_{1} = xy$.  
By substituting $\Lambda_{1} = xy$ into Eq.(69) and performing a detailed expansion, the corresponding conserved vectors are obtained as follows:
\begin{equation}
	\begin{aligned}
		\Phi^{t}_{1} &= x y\, u_{x}, \\[6pt]
		\Phi^{x}_{1} &= \gamma x y\, u_{xxx} - \gamma y\, u_{xx} + e x y\, u_{x} 
		- e y\, u + \alpha x y\, u\, u_{x} - \tfrac{1}{2}\alpha y\, u^{2}, \\[6pt]
		\Phi^{y}_{1} &= d x y\, u_{y} - d x\, u.
	\end{aligned}
\end{equation}
\textbf{Case 2: Conservation law corresponding to the multiplier $\Lambda_{2} = y\, c_{2}(t)$} 
\newline 
For the second case, the multiplier is chosen as $\Lambda_{2} = y\, c_{2}(t)$.  
Substituting $\Lambda_{2}$ into Eq.(69) and simplifying, the associated conserved vectors are determined to be
\begin{equation}
	\begin{aligned}
		\Phi^{t}_{2} &= 0, \\[6pt]
		\Phi^{x}_{2} &= y\, c_{2}(t) \left[ 
		\alpha\, u\, u_{x} 
		+ e\, u_{x} 
		+ \gamma\, u_{xxx} 
		+ u_{t} 
		\right], \\[6pt]
		\Phi^{y}_{2} &= d\, y\, c_{2}(t)\, u_{y} - d\, c_{2}(t)\, u.
	\end{aligned}
\end{equation}
\textbf{Case 3: Conservation law corresponding to the multiplier $\Lambda_{3} = c_{3}(t)$}  
\newline
In this case, the multiplier is taken as $\Lambda_{3} = c_{3}(t)$.  
Expanding Eq.(69) with this multiplier yields the conserved vector components in the form
\begin{equation}
	\begin{aligned}
		\Phi^{t}_{3} &= 0, \\[6pt]
		\Phi^{x}_{3} &= c_{3}(t) \left[ 
		\alpha\, u\, u_{x} 
		+ e\, u_{x} 
		+ \gamma\, u_{xxx} 
		+ u_{t} 
		\right], \\[6pt]
		\Phi^{y}_{3} &= d\, c_{3}(t)\, u_{y}.
	\end{aligned}
\end{equation}
\textbf{Case 4: Conservation law corresponding to the multiplier $\Lambda_{4} = x\, c_{1}(t)-\dfrac{y^{2} c'_{1}(t)}{2 d}$} 
\newline 
For the fourth multiplier, we set $\Lambda_{4} = x\, c_{1}(t)-\dfrac{y^{2} c'_{1}(t)}{2 d}$.  
Substituting this multiplier into Eq.(69) and performing the necessary computations, the conserved vector components are found to be
\begin{equation}
	\begin{aligned}
		\Phi^{t}_{4} &= -\, c_{1}(t)\, u, \\[6pt]
		\Phi^{x}_{4} &= x\, c_{1}(t)\, \gamma\, u_{xxx}+x\, c_{1}(t)\, u_{t}
		- \gamma\, c_{1}(t)\, u_{xx} 
		+ \alpha\, x\, c_{1}(t)\, u\, u_{x} 
		+ e\, x\, c_{1}(t)\, u_{x} 
		- e\, c_{1}(t)\, u \\[4pt]
		&\quad - \frac{\alpha\, y^{2}\, c'_{1}(t)\, u\, u_{x}}{2d}
		- \frac{y^{2}\, c'_{1}(t)\, e\, u_{x}}{2d}
		- \frac{y^{2}\, c'_{1}(t)\, \gamma\, u_{xxx}}{2d}
		- \frac{y^{2}\, c'_{1}(t)\, u_{t}}{2d}
		- \frac{\alpha\, c_{1}(t)\, u^{2}}{2}, \\[6pt]
		\Phi^{y}_{4} &= -\frac{1}{2} y^{2} c'_{1}(t)\, u_{y} 
		+ c'_{1}(t)\, y\, u 
		+ x\, c_{1}(t)\, d\, u_{y}.
	\end{aligned}
\end{equation}
\section{Conclusion}
 In this paper, we have analysed dimensional reduced (1+2)-dimensional Kudryashov-Sinelshchikov equation which represents the fourth order nonlinear PDE describing nonlinear wave propagation in bubbly liquid  with viscosity and heat transfer effects using Lie point symmetry, Painlevé analysis, and conservation laws using multiplier approach. We obtain infinite dimensional Lie algebra due to the presence of arbitrary functions $a(t)$, $g(t)$, and $h(t)$. By using commutation product property between the vector fields we obtain system of ODEs, manuanlly solving these system give rise to relation betweeen these arbitrary functions. By choosing appropriate functional forms for this arbitrary functions, led to the reduction of the original fourth-order nonlinear (1+2)-dimensional PDE into a fourth-order nonlinear (1+1)-dimensional PDE. Again, applying the symmetry analysis to the symmetry-reduced PDE we obtain symmetry-reduced fourth order nonlinear ODEs. With the help of Painlev\'e analysis, we obtain the solution of reduced ODEs in Laurent series from. We can see that two of the reduced ODEs passes Painlev\'e test and admit
Laurent series solutions, whereas one equation fails the test, indicating non-integrability. 
\newline
It is observed that the compatibility conditions at the resonance levels are satisfied for certain parameter choices, namely \(d = 0\) and \(\gamma = 0\) with \(\alpha \neq 0\). Under these conditions, the equation passes the Painlev\'e test and admits solutions in the form of a Laurent series. However, these restrictions simplify the original model: when \(d = 0\), the equation reduces to a \((1+1)\)-dimensional partial differential equation, whereas for \(\gamma = 0\), the highest-order term in the equation vanishes. 
\newline
With the help of multiplier approach, we find the conservation laws for governing nonlinear partial differential equation. We obtain an infinite numbers of multipliers due to the presence of arbitrary functions $c_{1}(t), c_{2}(t)$ and $c_{3}(t)$. Hence, We get four cases and the Conservation laws were verified for all the cases. 

The precise solutions that were found in this work offer a great deal of insight into the physical characteristics of pressure waves in bubbly liquids showing how nonlinear effects and viscosity, and heat transfer, affect the amplitude of waves, wave velocity and wave stability. The applications of these findings are practically relevant in a variety of applications: cavitation in hydraulic systems, acoustics in underwater, biomedical ultrasound, and oil-gas transport, where the bubble-liquid interaction largely influences the performance of these systems. Moreover, the obtained analytical solutions can be used as credible standards to validate numerical simulations as well as experimental results of nonlinear wave propagation in complex fluids. In general, this work contributes to the theoretical and practical knowledge on nonlinear dispersive-dissipative systems and provides a systematic framework of analysis of other higher-dimensional nonlinear evolution equations.

\section*{Acknowledgements}
RS and AP acknowledges the project funding support of Anusandhan National Research Foundation (ANRF), India under the Core Research Grant with File No: CRG/2023/005418, dated 20 August, 2024.
\newline
This research is partially supported by the DST-FIST grant SR/FST/MS-I/2024/173 to the Department of Mathematics, Pondicherry University,
Puducherry, India.
\section*{Conflict of Interest}
The authors declare that there are no competing interests among the authors of this work.

\end{document}